\documentclass[aps,pre,twocolumn,superscriptaddress]{revtex4-1}

\usepackage{amsmath,amssymb}
\usepackage{graphicx}

\begin{document}


\title{Tensor Renormalization Group with Randomized Singular Value Decomposition}


\author{Satoshi Morita}
\email[]{morita@issp.u-tokyo.ac.jp}
\affiliation{The Institute for Solid State Physics, The University of Tokyo,
 Kashiwa, Chiba 277-8581, Japan}

\author{Ryo Igarashi}
\affiliation{Information Technology Center, The University of Tokyo,
 Bunkyo-ku, Tokyo 113-8658, Japan}

\author{Hui-Hai Zhao}
\affiliation{RIKEN Brain Science Institute, Wako-shi, Saitama 351-0198, Japan}

\author{Naoki Kawashima}
\affiliation{The Institute for Solid State Physics, The University of Tokyo,
 Kashiwa, Chiba 277-8581, Japan}



\date{\today}

\begin{abstract}
An algorithm of the tensor renormalization group is proposed based on a
randomized algorithm for singular value decomposition. Our algorithm is
applicable to a broad range of two-dimensional classical models.  In the
case of a square lattice, its computational complexity and memory usage
are proportional to the fifth and the third power of the bond dimension,
respectively, whereas those of the conventional implementation are of
the sixth and the fourth power.  The oversampling parameter larger than
the bond dimension is sufficient to reproduce the same result as full
singular value decomposition even at the critical point of the
two-dimensional Ising model.
\end{abstract}


\maketitle


\section{Introduction}

Tensor networks are becoming powerful tools in the study of strongly
correlated condensed matter
physics~\cite{Cirac_JPhysA2009,Orus_AnnPhys2014}.  A classical example
is the density matrix renormalization
group~\cite{DMRG1992,fannes1992,Ostlund1995}, which can be viewed as a
variational method based on a one-dimensional tensor network, i.e., the
matrix product state. Its higher-dimensional generalization, such as the
projected entangled pair state (PEPS)~\cite{PEPS} and projected
entangled simplex state (PESS)~\cite{PESS2014}, is quite successful.
In classical systems, the partition functions can be expressed
as tensor networks~\cite{Baxter, SRGfull}, so that the physical
properties of the systems can be obtained by the contraction of tensor
networks.

One of the main goals of developing tensor network algorithms is to find
efficient and accurate methods for contracting tensor
networks. Real-space renormalization by coarse-graining tensor networks,
including the tensor renormalization group (TRG)~\cite{TRG} method and its
derivatives~\cite{SRG,SRGfull,SRGfinite,HOTRG,TNR,loop-TNR}, is
an efficient numerical method for the contraction of tensor networks.  However
these methods require huge computational time and memory usage even
though they are polynomially proportional to the system size.  Thus
complexity reduction without loss of accuracy is desired.

Decomposition and contraction are major parts of most tensor
network methods.  The former splits a tensor into two tensors.  In
general, exact decomposition requires a huge computational cost and
memory because of the large bond dimension between the two tensors.  To avoid
this problem, an approximation based on singular value decomposition
(SVD) is often used.  One can keep the bond dimension finite by
truncating small singular values.  Therefore, what is really
necessary in most cases is a partial SVD, rather than a full
SVD. Actually, however, truncation after a full SVD is a frequently
used procedure, despite it having a different (worse) computational
complexity than the partial SVD.  This is partially due to the lack of
efficient and easy-to-use libraries supporting partial SVD on the latest
parallel machines.

One of the partial SVD algorithms is the Arnoldi method, which is an
iterative algorithm based on the Krylov subspace~\cite{Arnoldi}. To
create the Krylov subspace, this method iterates matrix-vector
products. However,  a matrix-vector product is generally less efficient
than a matrix-matrix product because the memory band-width becomes narrow
on the latest massively parallel machines.

Recently, a partial SVD algorithm based on the low-rank approximation
using a randomized algorithm was proposed, which was called randomized
singular value decomposition (RSVD)~\cite{RSVD}.  To obtain a projector
to the subspace spanned by singular vectors corresponding to leading
singular values, a random matrix is multiplied to a target matrix to be
decomposed.
The computational efficiency of matrix-matrix products is the advantage of
RSVD over the Arnoldi method, although their computational costs are of the
same order.  In Refs. \cite{TEBD_RSVD,TEBD_RSVD2}, the RSVD was applied to the
time-evolving block-decimation (TEBD)~\cite{TEBD2003,TEBD2004} method
based on a matrix product state, which is a one-dimensional tensor
network, and its speed-up compared with full SVD was confirmed.
However, this method does not reduce computational
complexity with respect to the matrix size.

In this paper, we apply RSVD to a two-dimensional tensor network and
investigate its efficiency and accuracy.  Especially, we focus on the TRG
method~\cite{TRG}, one of the simplest real-space renormalization
schemes, and propose a scheme of TRG using RSVD.  Its computational
complexity scales as $O(\chi^5)$ with the bond dimension $\chi$, while
the original TRG method is $O(\chi^6)$.  Although using the partial
SVD is vital in reducing the order of the complexity, SVD is not the
only part that yields the $\chi^6$ dependence. Therefore, as we discuss
below, we need an alternative scheme for the whole procedure of TRG to reduce the
order.  Its memory usage is also reduced from $O(\chi^4)$ to
$O(\chi^3)$.

This paper is organized as follows.  In the next section, we propose our
scheme of TRG with RSVD.  We also briefly review the original TRG
method and the RSVD algorithm.  In the third section, we report benchmark
results of our method on the two-dimensional Ising model.  We show
the scaling of computational time and dependence of its accuracy on the
oversampling parameter for RSVD.  The performance of the power iteration
scheme is also investigated.  The last section is devoted to the
summary.

\section{Algorithms}
\subsection{Tensor Renormalization Group}

First, we review the TRG method for a translation invariant tensor
network on a square lattice.  Let us consider that the
local tensors $T^{(0)}$ are located on each lattice site. A contraction
of all local tensors gives the partition function as
\begin{equation}
 Z = \text{Tr}\, \prod_{i} T_{x_iy_ix_i'y_i'}^{(0)},
\end{equation}
where $i$ runs over all lattice sites, and the operation $\text{Tr}$ is
to sum over all the tensor indices.  By redefining a lattice site,
adding auxiliary degrees of freedom, and/or taking a local summation,
various short-range interaction models on a two-dimensional periodic
lattice can be cast into a nearest-neighbor-interaction model on a square
lattice, for which the tensor can be expressed, in general, as
\begin{equation}
 T_{xyx'y'}^{(0)} = \sum_{s}W_{sx}W_{sy}W_{sx'}^*W_{sy'}^*,
  \label{eq:initial_tensor}
\end{equation}
where $W$ is the square root of the local Boltzmann factor,
\begin{equation}
 \sum_{x}W_{sx} W_{s'x}^* =
  \exp\left(-\beta h_{ss'}\right).
\end{equation}
Here, $\beta=1/T$ is the inverse temperature and $h_{ss'}$ denotes the local
Hamiltonian.
Classical models with continuous degrees of freedom can also be
represented as a finite-dimension tensor network with high accuracy \cite{HOTRG_XY}.

\begin{figure}
 \includegraphics[scale=0.25]{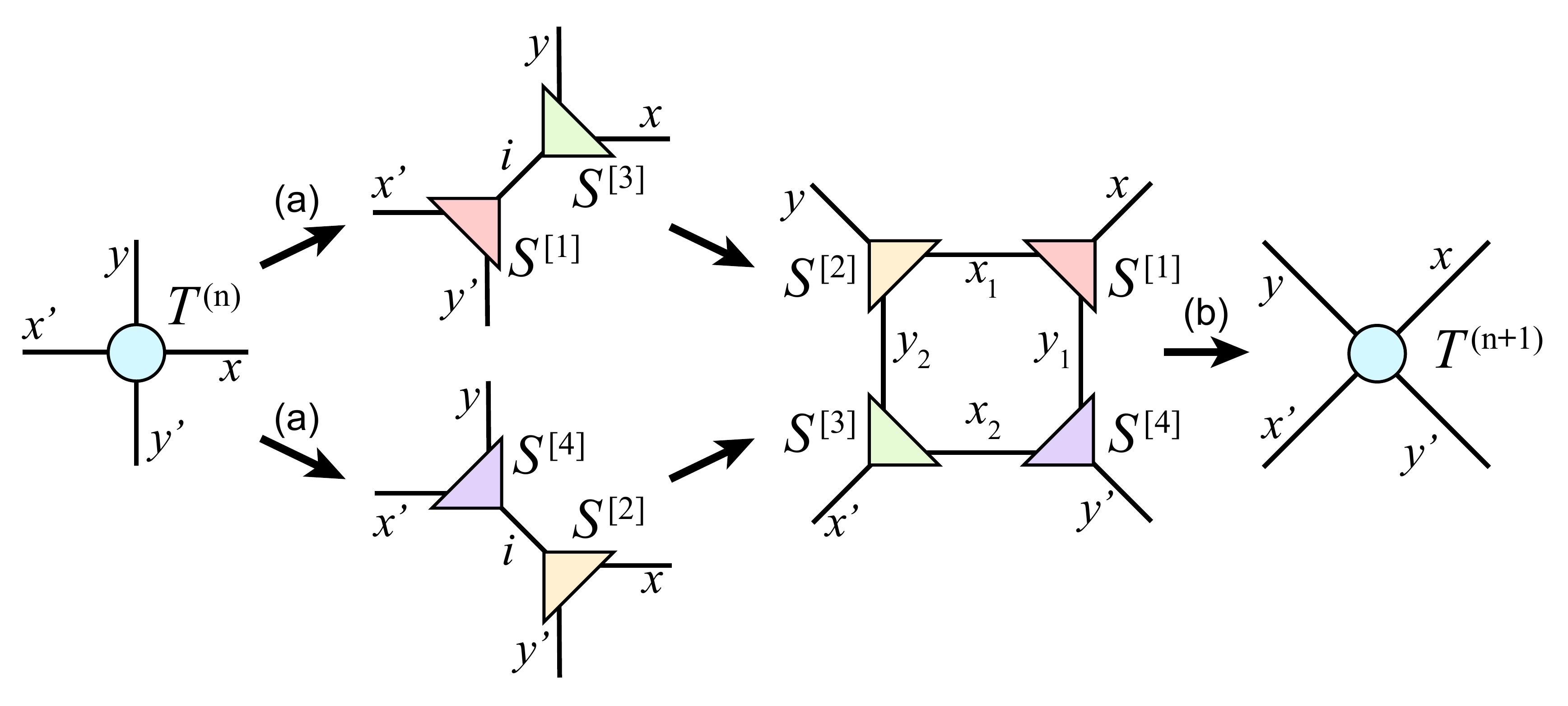}
 \caption{\label{fig:trg} (Color online) Graph representation of the TRG
 algorithm. (a) The local tensor is decomposed into two third-order
 tensors in two ways. (b) Contraction of four third-order tensors
 provides the renormalized tensor.}
\end{figure}

The TRG method consists of two key steps, decomposition and contraction.
In the first step of TRG, the local tensor is approximated by the product of two
third-order tensors in two ways, as shown in Fig.~\ref{fig:trg}(a),
\begin{gather}
 T_{xyx'y'}^{(n)} \simeq \sum_{i=1}^{\chi}
  S_{xy,i}^{[3]} S_{x'y',i}^{[1]}\\
 T_{xyx'y'}^{(n)} \simeq \sum_{i=1}^{\chi}
  S_{xy',i}^{[2]} S_{x'y,i}^{[4]}
\end{gather}
where $\chi$ denotes the maximum bond dimension which determines the accuracy
of the algorithm.
The truncation based on the singular value decomposition,
 $T_{xy,x'y'}=\sum_{i} s_i U_{xy,i} V^*_{x'y',i}$ provides minimum error
defined by the Frobenius norm. We assume that the singular values ${s_i}$
satisfy $s_1\geq s_2\geq \cdots$. The decomposed tensors $S^{[1]}$ and
$S^{[3]}$ are calculated as
\begin{gather}
 S_{xy,i}^{[3]} = \sum_{i=1}^{\chi} \sqrt{s_i} U_{xy,i}\\
 S_{x'y',i}^{[1]} = \sum_{i=1}^{\chi} \sqrt{s_i} V^*_{x'y',i}.
\end{gather}
The other tensors, $S^{[2]}$ and $S^{[4]}$, are obtained by SVD of a matrix
$T'_{xy',x'y}=T_{xyx'y'}$.
In the second step, we calculate the renormalized tensor by the contraction of
four third-order tensors, as shown in Fig.~\ref{fig:trg}(b),
\begin{equation}
 T_{xyx'y'}^{(n+1)} = \sum_{x_1 x_2 y_1 y_2}
  S_{x_1y_1,x}^{[1]}S_{x_1y_2,y}^{[2]}
  S_{x_2y_2,x'}^{[3]}S_{x_2y_1,y'}^{[4]}.\label{eq:contract_4}
\end{equation}
The resulting tensor network tilts by 45 degrees and the lattice spacing
increases by a factor of $\sqrt{2}$.

The computational cost to obtain all the singular values and vectors
scales as $O\left(\chi^6\right)$, while the partial SVD takes
$O(\chi^5)$ cost.  We note that the computational cost of contraction in
Eq.(\ref{eq:contract_4}) also scales as $O\left(\chi^6\right)$.  Thus,
we need to reduce both the computational costs of tensor decomposition
and construction.  The memory usage of the original TRG algorithm scales
as $O(\chi^4)$.

\subsection{Randomized algorithm for SVD}

In this subsection, we briefly review the randomized algorithm for
singular value decomposition (RSVD)~\cite{RSVD}.  Let us consider an
$m\times n$ matrix $A$ to be decomposed. The goal is to obtain the
leading $k$ singular values and corresponding singular vectors of $A$.

The first stage of RSVD is to obtain the low-rank
approximation of $A$ as
\begin{equation}
 A \simeq QQ^\dagger A.\label{eq:low_rank_approx}
\end{equation}
Here the basis matrix $Q$ is an $m\times (k+p)$ matrix whose columns are
orthogonal, i.e. $Q^\dagger Q$ is the identity matrix. We introduce the
oversampling parameter $p$ which determines the accuracy of RSVD.  The
optimal solution of $Q$ that minimizes the Frobenius distance
$\|A-QQ^\dagger A \|$ is given by the matrix whose columns are the left
singular vectors corresponding to the leading $(k+p)$ singular values.

To obtain the basis matrix $Q$, we use an $n\times (k+p)$ random matrix
$\Omega$. Reorthogonalization of an $m\times (k+p)$ matrix $Y\equiv
A\Omega$ by QR decomposition ($Y=QR$) or the Schmidt
orthogonalization provides the matrix $Q$.
The columns of a random matrix $\Omega$ will be
linearly independent with high probability.  If the rank of $A$ is
$(k+p)$, the columns of $Y$ will span the image of the linear
transformation induced by $A$. Thus, the reorthogonalization of $Y$
produces the orthogonal basis for the image of $A$.

In this paper, we use the standard Gaussian matrix as $\Omega$,
whose components are independently drawn from the normal distribution.
However, a choice of the random distribution is not essential for the
accuracy of RSVD.  We confirmed that the uniform distribution produced
almost the same results as the Gaussian distribution.  Note that the
elements of $\Omega$ could be complex when $A$ has complex entries.

In the second stage of RSVD, we form the $(k+p)\times n$ matrix $B\equiv
Q^\dagger A$ and compute the full SVD of $B$.  By dropping smaller singular
values, we obtain $B \simeq \tilde{U} \Sigma V^\dagger$, where
$\tilde{U}$ and $V$ are $(k+p)\times k$ and $n\times k$ matrices with
orthonormal columns, respectively. The $k\times k$ diagonal matrix
$\Sigma$ holds the largest $k$ singular values of $B$. The columns
of matrix $V$ approximate the right singular vectors of $A$. Finally, we
form the $m\times k$ matrix $U=Q\tilde{U}$ with the left singular
vectors of $A$.

The upper bound of expectation error of the low-rank approximation
Eq.(\ref{eq:low_rank_approx}) is estimated analytically as
\begin{equation}
 \left\langle
  \|A-QQ^\dagger A\|
 \right\rangle
 \leq \left(1+\frac{k}{p-1}\right)^{1/2}
 \left(\sum_{j>k} s_j^2\right)^{1/2},
\end{equation}
where the angle brackets stand for expectation with respect to the Gaussian
test matrix $\Omega$~\cite{RSVD}.  The optimal solution of the $k$-rank
approximation obtained by SVD has the minimum Frobenius-norm error
$(\sum_{j>k} s_j^2)^{1/2}$.  If the singular values decay
exponentially or faster as a function of the index, the randomized
algorithm provides accurate decomposition with small $p$ and its error
is of order $s_{k+1}$.

The power iteration scheme improves the accuracy of low-rank
approximation (\ref{eq:low_rank_approx}), in which
$Y=A\Omega$ is replaced
into $Y_{2q+1}=(AA^\dagger)^q A\Omega$ or $Y_{2q}=(AA^\dagger)^{q}
\Omega'$.  Here, $\Omega'$ is an $m\times(k+p)$ random matrix.  Clearly,
the upper bound of expectation error for $Y_r$ is proportional to
$(\sum_{j>k} s_j^{2r})^{1/2}$.  Thus the power iteration reduces
the approximation error exponentially with the power $r$ while the
computational cost is proportional to $r$.  The following algorithm,
which is algebraically equivalent to the power iteration, is useful in practice
to reduce the rounding error in floating-point arithmetic.  First,
form $Y_1=A\Omega$ and compute its QR decomposition $Y_1=Q_1 R_1$.
Next, repeat $2q$ times the matrix-matrix products and the QR
decompositions,
\begin{gather*}
 Y_{2j}=A^\dagger Q_{2j-1}=Q_{2j} R_{2j}\\
 Y_{2j+1}=AQ_{2j}^\dagger=Q_{2j+1} R_{2j+1}.
\end{gather*}
The resulting basis matrix $Q_{2q+1}$ is the same as the QR decomposition
of $Y_{2q+1}$.  In the case of $r=2q$, we start from the QR
decomposition of $A^\dagger \Omega'$.

In the case of $(k+p) < m,n$, the computational cost of RSVD is
$O(mn(k+p))$ which comes from the matrix-matrix products $A\Omega$ and
$Q^\dagger A$. The QR decomposition of $Y$ and the full SVD of the
matrix $B$ have a smaller cost than either one of the matrix
multiplications.  If the oversampling parameter $p$ is less than $O(k)$,
the cost of RSVD is $O(mnk)$, which is the same as that of the Arnoldi
method.
The advantage of RSVD over the Arnoldi method is that a
matrix-matrix product is much more efficient than a matrix-vector
product because the performance of a matrix-vector product is often
limited by the memory bandwidth.

In the TRG algorithm on the square lattice, the local tensor $T$ is
transformed into a $\chi^2\times\chi^2$ matrix and truncated by keeping
leading $\chi$ singular values, i.e., $m=n=\chi^2$ and
$k=\chi$. Therefore, the computational cost of tensor decomposition
with RSVD is $O(\chi^5)$ if the oversampling parameter $p$ is at most of
order $\chi$.
If we utilize the power iteration scheme, the order of the computational
cost increases only by a factor $r$.

\subsection{$O(\chi^5)$ algorithm of TRG}

\begin{figure}
 \includegraphics[scale=0.25]{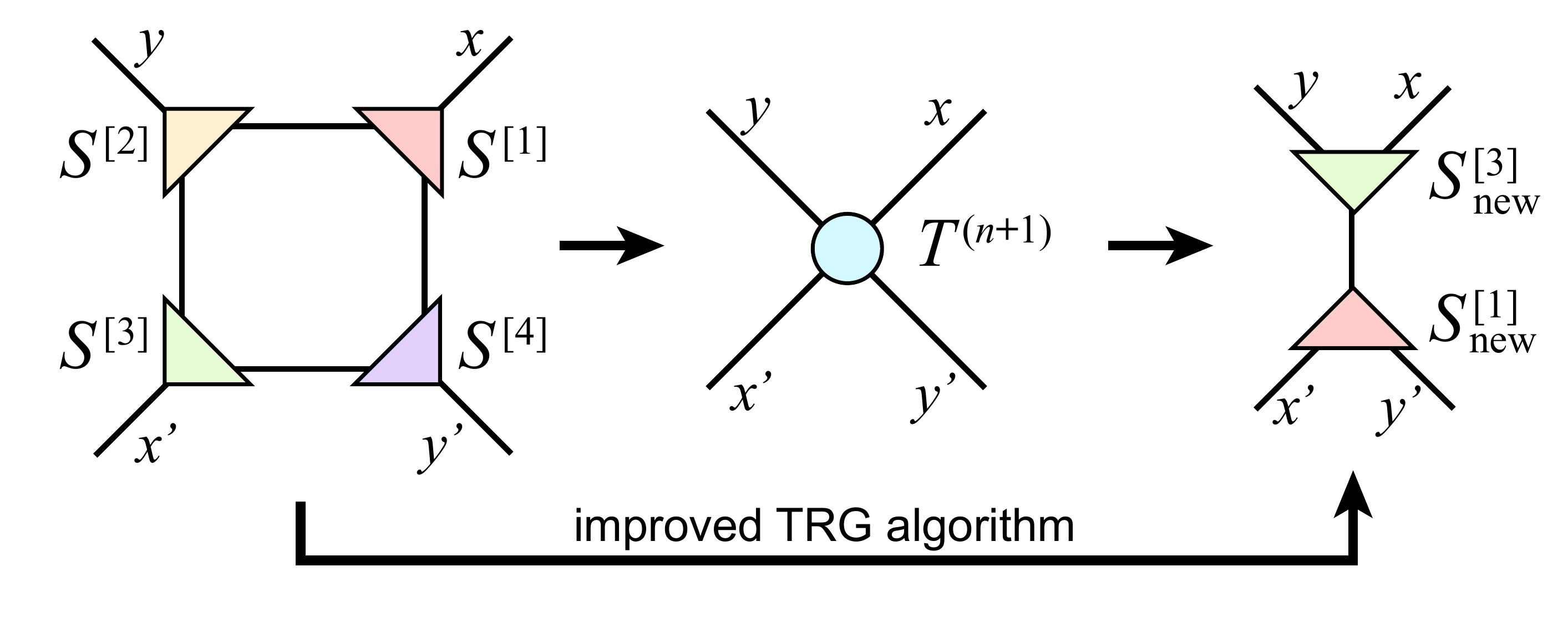}

 \caption{\label{fig:trg2} (Color online) The chain of deformation in
 the TRG algorithm.  Our improved algorithm skips the intermediate step
 of calculating the fourth-order tensor $T$.}
\end{figure}

While the cost of tensor decomposition is reduced to $O(\chi^5)$ by
using partial SVD, the total cost of TRG is still $O(\chi^6)$ owing to
tensor contraction in Eq.(\ref{eq:contract_4}).
The present section shows that we can reduce the total cost down to
$O(\chi^5)$ by working directly with the four third-order tensors
$S^{[i]}$ without actually computing the fourth-order tensor $T$. In
other words, the iterative SVD techniques such as RSVD make it possible
to skip the intermediate step of computing $T$ in the chain of
deformation as shown Fig.~\ref{fig:trg2}.
The key observation is that in the procedure in RSVD
described in the previous section, we actually do not need the matrix
elements as long as we can compute the results of the matrix operation
on an arbitrary vector or matrix. In the present case, we can operate the
four $S$ tensors one-by-one on a given tensor to obtain the same
result as operating $T$ on it. Therefore, we do not need the explicit
form of the tensor $T$.
Moreover, our improved algorithm reduces the memory usage from
$O(\chi^4)$ to $O(\chi^3)$.

\begin{figure}
 \includegraphics[scale=0.25]{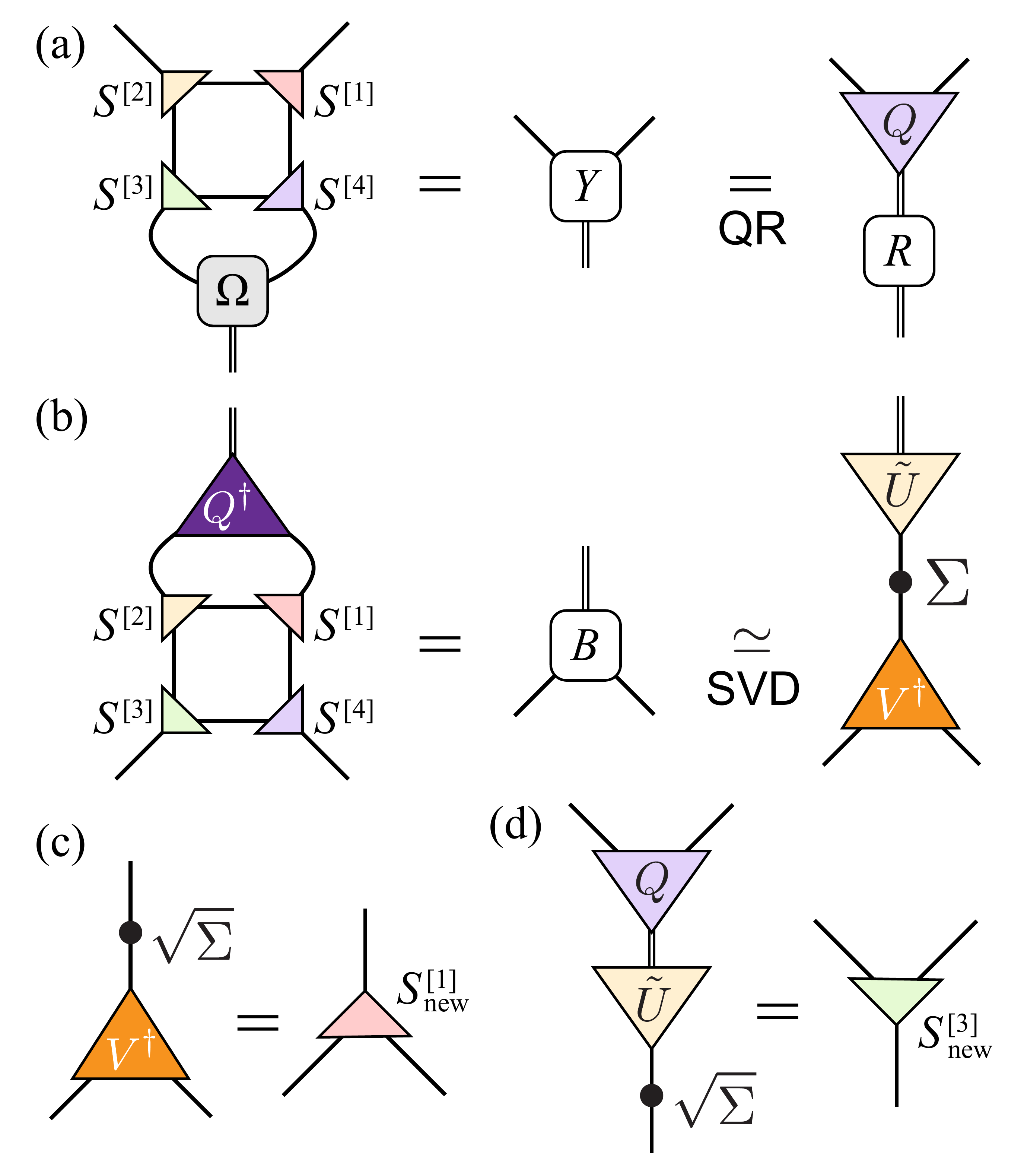}
 \caption{\label{fig:trg_chi5} (Color online) The improved TRG algorithm
 with $O(\chi^5)$ computational cost.  The double lines indicate bonds
 with dimension $\chi+p$.  (a) The first stage of RSVD.  (b) The second
 stage of RSVD.  (c, d) Updated third-order tensors $S^{[1]}_\text{new}$
 and $S^{[3]}_\text{new}$.}
\end{figure}

The graphic representation of the improved TRG algorithm is shown in
Fig.~\ref{fig:trg_chi5}. The solid bonds have dimension $\chi$, while the
double lines have dimension $\chi+p$ for the oversampling of RSVD.  This
figure shows how to generate $S^{[1]}_{\text{new}}$ and
$S^{[3]}_{\text{new}}$ from four tensors $\{S^{[i]}\}$. The other
tensors $S^{[2]}_{\text{new}}$ and $S^{[4]}_{\text{new}}$ can be
obtained by connecting the random tensor $\Omega$ to $S^{[2]}$ and
$S^{[3]}$.  The initial tensors of $S^{[i]}$ are straightforwardly
obtained from Eq.(\ref{eq:initial_tensor}). For example,
$S^{[1]}_{x'y',s}=W_{sx'}W_{sy'}$ and $S^{[3]}_{xy,s}=W_{sx}W_{sy}$.

The key diagrams in this algorithm are Figs.~\ref{fig:trg_chi5}(a) and
\ref{fig:trg_chi5}(b) corresponding to matrix-matrix products
$Y=A\Omega$ and $B=Q^\dagger A$ in the previous section.  This
contraction of five third-order tensors is of order $\chi^5$ as long as
the oversampling parameter $p$ is less than or scaled as $\chi$.  As we
mentioned, the cost of contraction of the tensor network without
$\Omega$ and $Q^{\dagger}$ is $O(\chi^6)$.  The order of contractions is
important to reduce the computational cost~\cite{NCON}. For example, the
computational cost of $Y=S^{[1]}(S^{[2]}(S^{[3]}(S^{[4]}\Omega)))$
scales as $O(\chi^5)$, but $Y=(((S^{[1]}S^{[2]})S^{[3]})S^{[4]})\Omega$
scales as $O(\chi^6)$.

We also note that the loop blocking technique helps reduce the memory
usage of contractions. Some summation loops of indices are partitioned
into small blocks and then the summations over the blocks are postponed
after the other contractions.  In the case of Figs.~\ref{fig:trg_chi5}(a)
and \ref{fig:trg_chi5}(b), memory usage is reduced to $O(\chi^3)$ by applying this
technique to the index between $S^{[1]}$ and $S^{[4]}$ (see the details
in the Appendix).  We emphasize that this technique always reduces the
memory usage of intermediate tensors to at most the same order of the initial
and final tensor networks.

The power iteration scheme of RSVD is applicable to this algorithm
within the same order of computational cost and memory usage.  We can
use the similar diagrams of Fig.~\ref{fig:trg_chi5}.  For example, the QR
decomposition of $B$ $(=Y_2)$ instead of SVD yields the third-order
tensor $Q_2$ and the contraction of Fig.~\ref{fig:trg_chi5}(a) by
replacing $\Omega$ by $Q_2$ provides the third-order tensor
corresponding to $Y_3$.

\section{Numerical results}

To investigate the effect of randomness in RSVD and performance of the
improved TRG algorithm, we calculate the free energy of the Ising model
on the square lattice.  The initial tensor Eq.(\ref{eq:initial_tensor})
for an Ising model without an external magnetic field is given with a
$2\times 2$ matrix,
\begin{equation}
 W \equiv
  \begin{pmatrix}
   \sqrt{\cosh \beta J} & \sqrt{\sinh \beta J}  \\
   \sqrt{\cosh \beta J} & -\sqrt{\sinh \beta J}
  \end{pmatrix}.
\end{equation}
The critical temperature of this model is given by $\beta_c
J=\log(\sqrt{2}+1)/2$.

\begin{figure}
 \includegraphics[scale=0.45]{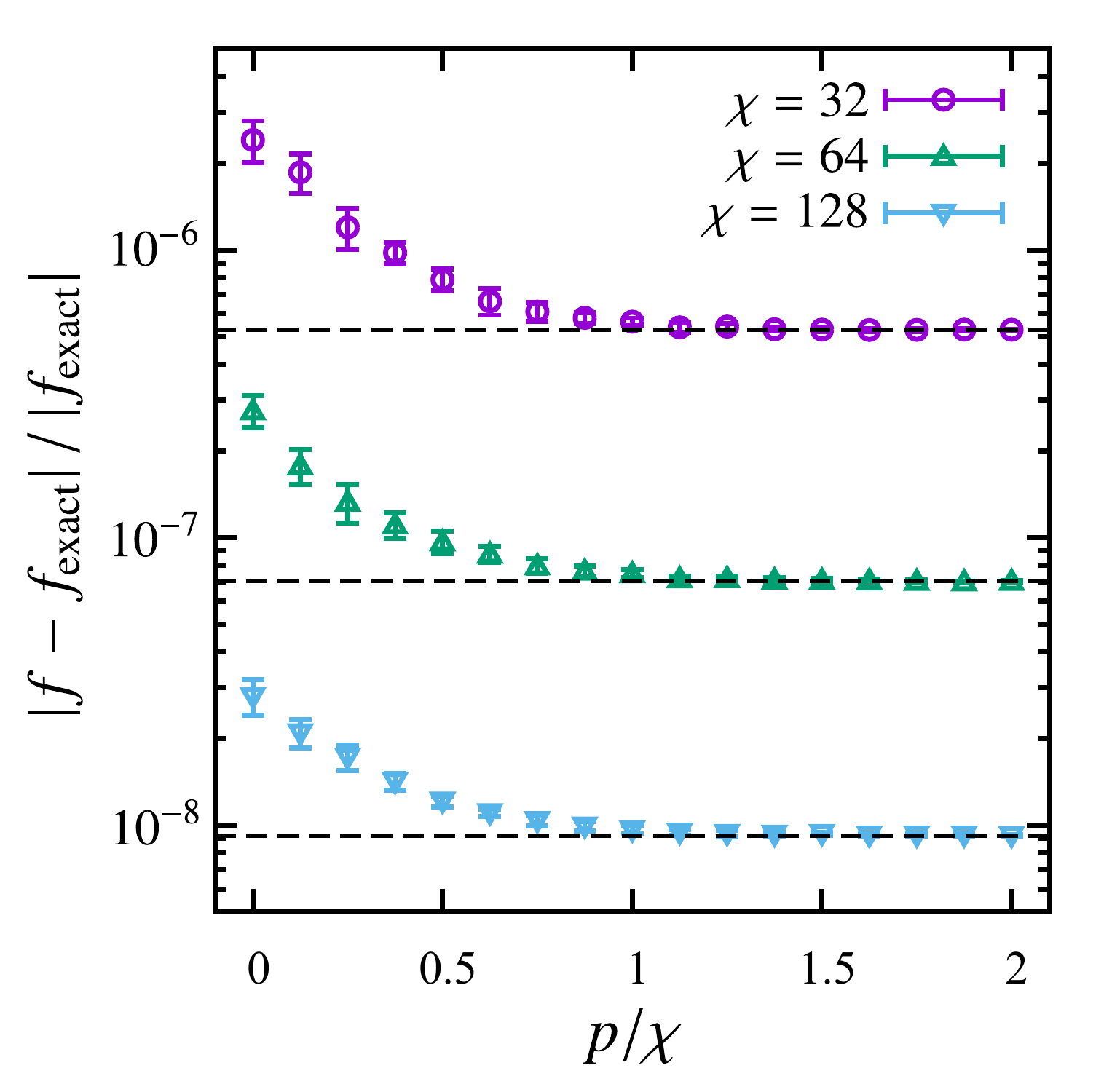}
 \caption{\label{fig:f_vs_oversamp} (Color online) Relative errors of
 the free energy at the critical temperature as a function of the
 oversampling parameter $p$. The error bars indicate the standard
 deviations estimated by more than 16 independent runs. The full-SVD
 results are shown by the horizontal dashed lines.}
\end{figure}

The relative errors of the free energy from the Onsager's solution
$f_\text{exact}$ at the critical temperature in the thermodynamic limit
are plotted against the oversampling parameter in
Fig.~\ref{fig:f_vs_oversamp}.  We iterated at least 36 TRG steps where
the renormalized tensor $T^{(36)}$ contains $2^{36}$ spins.  This TRG
step suffices for convergence of the free energy to the thermodynamic
limit at the critical temperature.  The error bars denote standard
deviations estimated by more than 16 independent runs.  The horizontal
dashed lines indicate results of the original algorithm using full SVD.
As expected, the improved TRG algorithm with a larger oversampling
parameter $p$ shows a smaller error and converges toward the full SVD
result.  Even at the critical temperature, $p\simeq \chi$ is sufficient
to provide the same results as full SVD independently of $\chi$.  In the
system away from the critical temperature, much smaller $p$ is
sufficient because of the rapid decay of the singular values.  The standard
deviation of free energy decreases with the oversampling parameter
because of the law of large numbers.

\begin{figure}
 \includegraphics[scale=0.44]{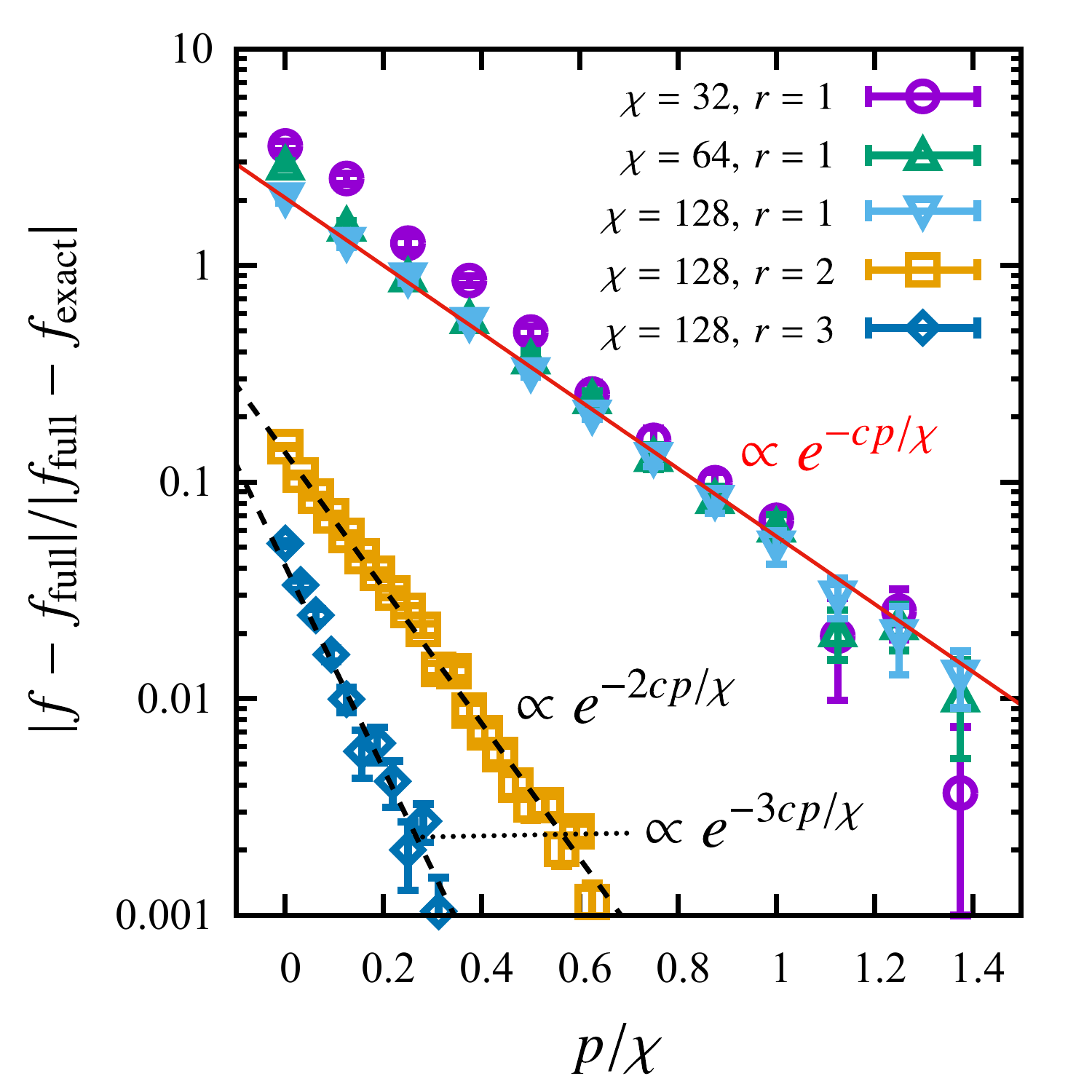}
 \caption{\label{fig:diff_from_full} (Color online) Difference in the
 free energy from the full SVD results decays exponentially with the
 oversampling parameter at the critical temperature.  The solid (red)
 line is obtained by fitting to the lower triangular data ($\chi=128$,
 $r=1$).  The same constant $c=3.60$ is used in the dashed (black)
 lines.}
\end{figure}

The accuracy of the power iteration scheme is shown in
Fig.~\ref{fig:diff_from_full}.  We found that the difference in the free
energy from the full SVD result exponentially decreases with the
oversampling parameter $p$ and the decay constant is proportional to the
number of power iterations $r$,
\begin{equation}
 |f-f_\text{full}| \propto   e^{-crp/\chi},
\end{equation}
This fact involves the upper bound of the error of the power iteration
scheme as mentioned before.  We estimated the coefficient $c=3.60$ at
the critical temperature with $\chi=128$.  The value of $c$ is nearly
independent of the bond dimension.  Since the power iteration scheme
enhances the decay of the singular values, a smaller value of the
oversampling parameter is sufficient for larger $r$.  For $r=2$,
$p\sim\chi/8$ achieves an accuracy comparable with $p=\chi$ without the
power iteration.  For $r=3$, even $p=0$ is sufficient.  From a viewpoint
of time to solution, however, the improved algorithm without the power
iteration is superior to the others.  For example, in the case of
$\chi=128$, the elapsed time per TRG step with $(r,p)=(1,128)$,
$(2,16)$, and $(3,0)$ is $85.2(1)$, $94.3(2)$, and $124.9(2)$ s, respectively.

\begin{figure}
 \includegraphics[scale=0.45]{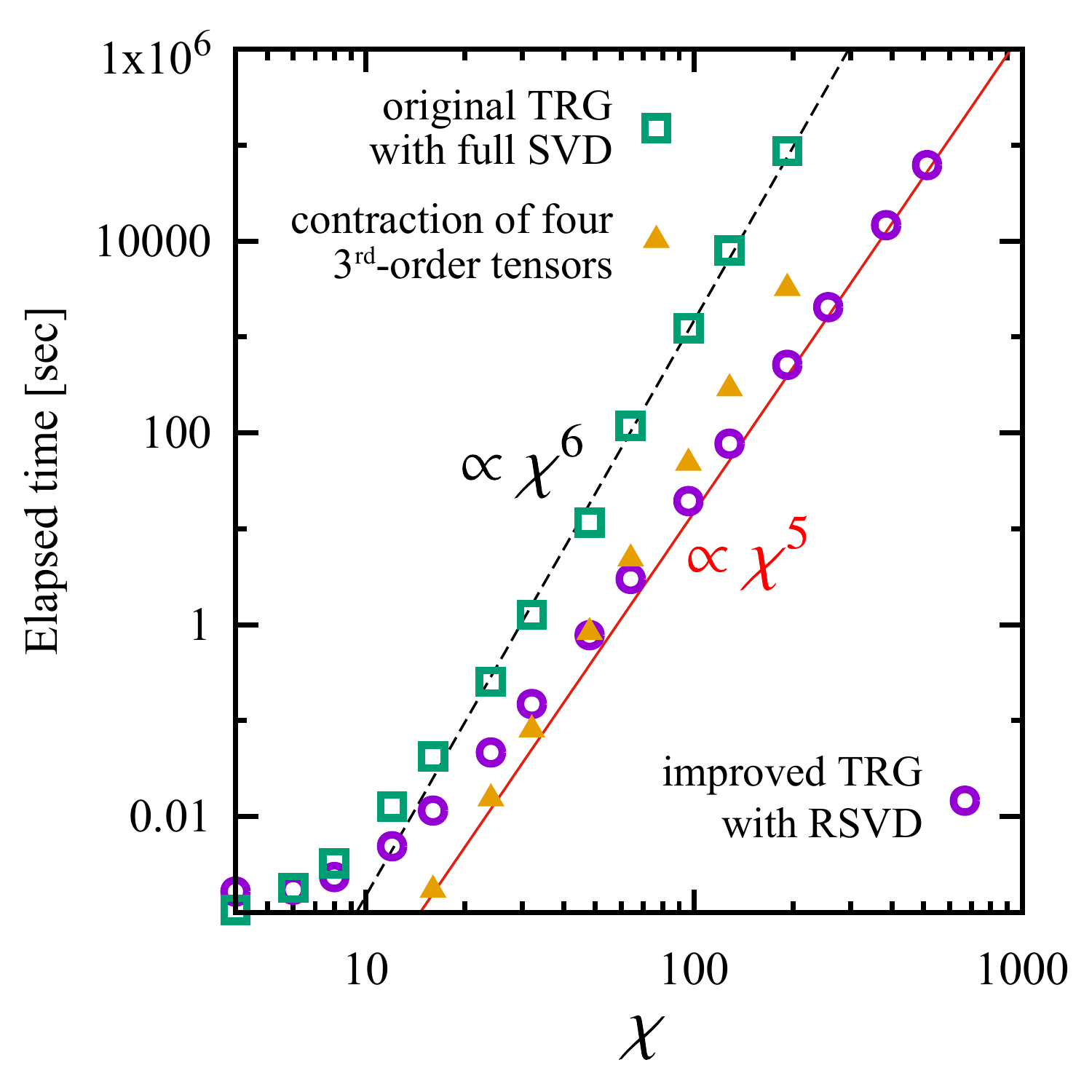}
 \caption{\label{fig:time_vs_chi} (Color online) Elapsed time per TRG
 step as a function of bond dimension $\chi$ is shown by open
 symbols. The improved TRG algorithm
 with RSVD (circles) scales as $O(\chi^5)$, while the original one with
 full SVD (squares) is $O(\chi^6)$.
 The oversampling parameter for RSVD is set as $p=\chi$ and the
 power iteration scheme is not used ($r=1$).
 The elapsed time of contraction
 Eq.(\ref{eq:contract_4}) in a step of TRG with full SVD is shown by
 the solid triangles.
 The solid (red) and dashed (black) lines, proportional to $\chi^5$ and
 $\chi^6$ respectively, are guides for the eyes.}
\end{figure}

The elapsed time per TRG step against bond dimension $\chi$ is plotted
in Fig.~\ref{fig:time_vs_chi}.  Here, we set the oversampling parameter of RSVD
as $p=\chi$ and do not use the power iteration scheme of RSVD ($r=1$).
The improved algorithm clearly follows
$\chi^5$ scaling, while the original one with full SVD scales as $O(\chi^6)$.
We achieved $\chi=512$ in the improved algorithm with
the aid of the loop blocking technique.
Although the most time-consuming part in the original algorithm is
full SVD, the contraction in Eq.~(\ref{eq:contract_4}) also scales as
$O(\chi^6)$, as shown in Fig.~\ref{fig:time_vs_chi}.  Thus, to achieve
the $\chi^5$ scaling, the replacement of full SVD with partial SVD is
insufficient and one needs to remove explicit construction of the fourth-order
tensor.
To compare the RSVD approach with other partial SVD methods with
$\chi^5$ scaling, we consider the Arnoldi method\cite{Arnoldi}.  It can
solve an SVD problem without explicit matrix or tensor construction and we
confirm that TRG with the Arnoldi method also shows $\chi^5$ scaling.
However we observe that the RSVD approach is around two times faster
than TRG with the Arnoldi method which takes $164.7(4)$ s per TRG
step for $\chi=128$.

We note that computational times were measured by simulations in a
single core on Intel Xeon E5-2697A (2.60 GHz) with 128 GB memory.  We
implemented the proposed TRG algorithm and original one by using the
script language Python. We used NumPy and SciPy~\cite{NUMPY, SCIPY}, the
fundamental packages for scientific computing with Python, for numerical
linear algebra.  These packages call LAPACK routines~\cite{LAPACK} for
full SVD and QR decomposition.  To compare with the present method based
on RSVD, we also used a partial SVD solver in the sparse linear algebra
module in SciPy, which is based on the implicitly restarted Arnoldi
method through ARPACK~\cite{ARPACK}.

\section{Conclusions}

In summary, we proposed a scheme of the TRG algorithm with $O(\chi^5)$
computational cost.  By using RSVD, we can avoid creating the
fourth-order tensor.  Numerical results on the two-dimensional Ising
model clearly show the $\chi^5$ scaling of computational time. Our
method is 100 times faster than the conventional method with full SVD
for $\chi=128$.  In addition, the memory usage scales as $O(\chi^3)$ by
using the loop-blocking technique.

The heaviest part in the RSVD algorithm is the matrix-matrix product.
The number of floating-point operations per memory access
(flops per byte, F/B) in the matrix-matrix product is proportional to
the linear size of the matrices.  On the other hand, the one for the
matrix-vector product is of order unity.  Since a narrow memory bandwidth
tends to be a bottleneck in current massively parallel machines, a larger F/B
is preferable.  Moreover, the matrix-matrix product can be accelerated
by general-purpose computing on graphics processing units and well
parallelized on distributed memory.  Therefore, RSVD is expected be more
efficient for a large matrix than the Krylov subspace methods including
the Arnoldi method.

Tensor decomposition by SVD commonly appears in other tensor network
methods and a tensor version of low-rank approximation
Eq.(\ref{eq:low_rank_approx}) is also a general and important
technique.
Although we applied RSVD only to the TRG method in this paper, it is
straightforward to use randomized algorithms instead of conventional
ones.
Thus, we believe that randomized algorithms would be useful to reduce
computational time and memory usage in many SVD-based tensor network
methods.

In the present paper, we have proposed improvements on the
tensor network computation by transforming the standard ``contraction
and decomposition'' procedure into multiplication among smaller tensors.
Here we emphasize that the proposed method reduced the computational
complexity of the whole procedure of the TRG method including not only
the SVD part but also the contraction.  Since these two components
dominate the computational time of most tensor network schemes, the
techniques presented in this paper would be useful in improving
most of the tensor network calculations in an essential way.

\begin{acknowledgments}
The authors would like to thank T~.Okubo, K~.Harada, and S.~Todo for
valuable discussions.  The computation in the present work is partially
executed on computers at the Supercomputer Center, ISSP, University of
Tokyo.  This research was supported by MEXT as "Exploratory Challenge on
Post-K computer" (Frontiers of Basic Science: Challenging the Limits),
by JSPS KAKENHI Grant No. 26730062, and by ImPACT Program of Council
for Science, Technology and Innovation (Cabinet Office, Government of
Japan).
\end{acknowledgments}

\appendix*
\section{Loop blocking technique}

The memory usage of contracting tensor networks can be reduced down to
the same order as the largest among the order of initial and final
tensors by using the loop blocking technique.  For example, let us consider
the contraction of Fig.~\ref{fig:trg_chi5}(a),
\begin{equation}
 Y_{xyz} = \sum_{\substack{x_1 y_1 x_2 y_2 \\x' y'}}
  S^{[1]}_{x_1y_1x}
  S^{[2]}_{x_1y_2y}
  S^{[3]}_{x_2y_2x'}
  S^{[4]}_{x_2y_1y'}
  \Omega_{x'y'z}.
\end{equation}
As we mentioned in the main text, the order of contractions,
$S^{[1]}(S^{[2]}(S^{[3]}(S^{[4]}\Omega)))$, achieves $O(\chi^5)$
computational cost.  However, some intermediate tensors such as
$S^{[4]}\Omega$ are fourth order. To avoid $O(\chi^4)$ memory usage, we
split the summation of the index $y_1$ between $S^{[1]}$ and $S^{[4]}$
into small blocks with a block size $\chi_b$.  Assuming the bond
dimension $\chi$ is divisible by the block size $\chi_b$ for
simplicity, the contraction with the loop blocking technique is precisely
represented as
\begin{align}
 Y_{xyz} = &\sum_{b=1}^{\chi/\chi_b} \Biggl[
 \sum_{x_1=1}^{\chi} \sum_{l=1}^{\chi_b}
 S^{[1]}_{x_1,\chi_b(b-1)+l,x} \notag \\
 &\times \Biggl(
 \sum_{y_2=1}^{\chi} S^{[2]}_{x_1y_2y}
 \Biggl(
 \sum_{x_2=1}^{\chi} \sum_{x'=1}^{\chi} S^{[3]}_{x_2y_2x'} \notag \\
 &\times \Biggl(
 \sum_{y'=1}^{\chi} S^{[4]}_{x_2,\chi_b(b-1)+l,y'}
 \Omega_{x'y'z}
 \Biggr)\Biggr)\Biggr) \Biggr].
\end{align}
Clearly, the memory usage of intermediate tensors is reduced to $O(\chi_b
\chi^3)$. Thus, it is $O(\chi^3)$ if the block size $\chi_b$ is of order
unity.
Our simulations in this paper typically used $\chi_b=8$, which reduced the
memory usage with $\chi=512$ from $550$ to $8.6$ GB.

Although splitting one loop is enough in this case, one needs to
block several loops in more complicate contractions. We note that the
loop blocking technique does not change the computational cost.

\bibliography{main}

\begin{thebibliography}{27}%
\makeatletter
\providecommand \@ifxundefined [1]{%
 \@ifx{#1\undefined}
}%
\providecommand \@ifnum [1]{%
 \ifnum #1\expandafter \@firstoftwo
 \else \expandafter \@secondoftwo
 \fi
}%
\providecommand \@ifx [1]{%
 \ifx #1\expandafter \@firstoftwo
 \else \expandafter \@secondoftwo
 \fi
}%
\providecommand \natexlab [1]{#1}%
\providecommand \enquote  [1]{``#1''}%
\providecommand \bibnamefont  [1]{#1}%
\providecommand \bibfnamefont [1]{#1}%
\providecommand \citenamefont [1]{#1}%
\providecommand \href@noop [0]{\@secondoftwo}%
\providecommand \href [0]{\begingroup \@sanitize@url \@href}%
\providecommand \@href[1]{\@@startlink{#1}\@@href}%
\providecommand \@@href[1]{\endgroup#1\@@endlink}%
\providecommand \@sanitize@url [0]{\catcode `\\12\catcode `\$12\catcode
  `\&12\catcode `\#12\catcode `\^12\catcode `\_12\catcode `\%12\relax}%
\providecommand \@@startlink[1]{}%
\providecommand \@@endlink[0]{}%
\providecommand \url  [0]{\begingroup\@sanitize@url \@url }%
\providecommand \@url [1]{\endgroup\@href {#1}{\urlprefix }}%
\providecommand \urlprefix  [0]{URL }%
\providecommand \Eprint [0]{\href }%
\providecommand \doibase [0]{http://dx.doi.org/}%
\providecommand \selectlanguage [0]{\@gobble}%
\providecommand \bibinfo  [0]{\@secondoftwo}%
\providecommand \bibfield  [0]{\@secondoftwo}%
\providecommand \translation [1]{[#1]}%
\providecommand \BibitemOpen [0]{}%
\providecommand \bibitemStop [0]{}%
\providecommand \bibitemNoStop [0]{.\EOS\space}%
\providecommand \EOS [0]{\spacefactor3000\relax}%
\providecommand \BibitemShut  [1]{\csname bibitem#1\endcsname}%
\let\auto@bib@innerbib\@empty
\bibitem [{\citenamefont {Cirac}\ and\ \citenamefont
  {Verstraete}(2009)}]{Cirac_JPhysA2009}%
  \BibitemOpen
  \bibfield  {author} {\bibinfo {author} {\bibfnamefont {J.~I.}\ \bibnamefont
  {Cirac}}\ and\ \bibinfo {author} {\bibfnamefont {F.}~\bibnamefont
  {Verstraete}},\ }\href {http://stacks.iop.org/1751-8121/42/i=50/a=504004}
  {\bibfield  {journal} {\bibinfo  {journal} {J. Phys. A: Math. Theor.}\
  }\textbf {\bibinfo {volume} {42}},\ \bibinfo {pages} {504004} (\bibinfo
  {year} {2009})}\BibitemShut {NoStop}%
\bibitem [{\citenamefont {Or\'us}(2014)}]{Orus_AnnPhys2014}%
  \BibitemOpen
  \bibfield  {author} {\bibinfo {author} {\bibfnamefont {R.}~\bibnamefont
  {Or\'us}},\ }\href@noop {} {\bibfield  {journal} {\bibinfo  {journal} {Ann.
  Phys.}\ }\textbf {\bibinfo {volume} {349}},\ \bibinfo {pages} {117} (\bibinfo
  {year} {2014})}\BibitemShut {NoStop}%
\bibitem [{\citenamefont {White}(1992)}]{DMRG1992}%
  \BibitemOpen
  \bibfield  {author} {\bibinfo {author} {\bibfnamefont {S.~R.}\ \bibnamefont
  {White}},\ }\href {\doibase 10.1103/PhysRevLett.69.2863} {\bibfield
  {journal} {\bibinfo  {journal} {Phys. Rev. Lett.}\ }\textbf {\bibinfo
  {volume} {69}},\ \bibinfo {pages} {2863} (\bibinfo {year}
  {1992})}\BibitemShut {NoStop}%
\bibitem [{\citenamefont {Fannes}\ \emph {et~al.}(1992)\citenamefont {Fannes},
  \citenamefont {Nachtergaele},\ and\ \citenamefont {Werner}}]{fannes1992}%
  \BibitemOpen
  \bibfield  {author} {\bibinfo {author} {\bibfnamefont {M.}~\bibnamefont
  {Fannes}}, \bibinfo {author} {\bibfnamefont {B.}~\bibnamefont
  {Nachtergaele}}, \ and\ \bibinfo {author} {\bibfnamefont {R.~F.}\
  \bibnamefont {Werner}},\ }\href
  {https://projecteuclid.org:443/euclid.cmp/1104249404} {\bibfield  {journal}
  {\bibinfo  {journal} {Commun. Math. Phys.}\ }\textbf {\bibinfo {volume}
  {144}},\ \bibinfo {pages} {443} (\bibinfo {year} {1992})}\BibitemShut
  {NoStop}%
\bibitem [{\citenamefont {\"Ostlund}\ and\ \citenamefont
  {Rommer}(1995)}]{Ostlund1995}%
  \BibitemOpen
  \bibfield  {author} {\bibinfo {author} {\bibfnamefont {S.}~\bibnamefont
  {\"Ostlund}}\ and\ \bibinfo {author} {\bibfnamefont {S.}~\bibnamefont
  {Rommer}},\ }\href {\doibase 10.1103/PhysRevLett.75.3537} {\bibfield
  {journal} {\bibinfo  {journal} {Phys. Rev. Lett.}\ }\textbf {\bibinfo
  {volume} {75}},\ \bibinfo {pages} {3537} (\bibinfo {year}
  {1995})}\BibitemShut {NoStop}%
\bibitem [{\citenamefont {Verstraete}\ and\ \citenamefont
  {Cirac}(2004)}]{PEPS}%
  \BibitemOpen
  \bibfield  {author} {\bibinfo {author} {\bibfnamefont {F.}~\bibnamefont
  {Verstraete}}\ and\ \bibinfo {author} {\bibfnamefont {J.~I.}\ \bibnamefont
  {Cirac}},\ }\href@noop {} {\bibfield  {journal} {\bibinfo  {journal} {arXiv}\
  } (\bibinfo {year} {2004})},\ \Eprint {http://arxiv.org/abs/cond-mat/0407066}
  {cond-mat/0407066} \BibitemShut {NoStop}%
\bibitem [{\citenamefont {Xie}\ \emph {et~al.}(2014)\citenamefont {Xie},
  \citenamefont {Chen}, \citenamefont {Yu}, \citenamefont {Kong}, \citenamefont
  {Normand},\ and\ \citenamefont {Xiang}}]{PESS2014}%
  \BibitemOpen
  \bibfield  {author} {\bibinfo {author} {\bibfnamefont {Z.~Y.}\ \bibnamefont
  {Xie}}, \bibinfo {author} {\bibfnamefont {J.}~\bibnamefont {Chen}}, \bibinfo
  {author} {\bibfnamefont {J.~F.}\ \bibnamefont {Yu}}, \bibinfo {author}
  {\bibfnamefont {X.}~\bibnamefont {Kong}}, \bibinfo {author} {\bibfnamefont
  {B.}~\bibnamefont {Normand}}, \ and\ \bibinfo {author} {\bibfnamefont
  {T.}~\bibnamefont {Xiang}},\ }\href {\doibase 10.1103/PhysRevX.4.011025}
  {\bibfield  {journal} {\bibinfo  {journal} {Phys. Rev. X}\ }\textbf {\bibinfo
  {volume} {4}},\ \bibinfo {pages} {011025} (\bibinfo {year}
  {2014})}\BibitemShut {NoStop}%
\bibitem [{\citenamefont {Baxter}(1982)}]{Baxter}%
  \BibitemOpen
  \bibfield  {author} {\bibinfo {author} {\bibfnamefont {R.~J.}\ \bibnamefont
  {Baxter}},\ }\href {\doibase 10.1137/1.9780898719628} {\emph {\bibinfo
  {title} {Exactly Solved Models in Statistical Mechanics}}}\ (\bibinfo
  {publisher} {Academic, London},\ \bibinfo {year} {1982})\BibitemShut
  {NoStop}%
\bibitem [{\citenamefont {Zhao}\ \emph {et~al.}(2010)\citenamefont {Zhao},
  \citenamefont {Xie}, \citenamefont {Chen}, \citenamefont {Wei}, \citenamefont
  {Cai},\ and\ \citenamefont {Xiang}}]{SRGfull}%
  \BibitemOpen
  \bibfield  {author} {\bibinfo {author} {\bibfnamefont {H.~H.}\ \bibnamefont
  {Zhao}}, \bibinfo {author} {\bibfnamefont {Z.~Y.}\ \bibnamefont {Xie}},
  \bibinfo {author} {\bibfnamefont {Q.~N.}\ \bibnamefont {Chen}}, \bibinfo
  {author} {\bibfnamefont {Z.~C.}\ \bibnamefont {Wei}}, \bibinfo {author}
  {\bibfnamefont {J.~W.}\ \bibnamefont {Cai}}, \ and\ \bibinfo {author}
  {\bibfnamefont {T.}~\bibnamefont {Xiang}},\ }\href {\doibase
  10.1103/PhysRevB.81.174411} {\bibfield  {journal} {\bibinfo  {journal} {Phys.
  Rev. B}\ }\textbf {\bibinfo {volume} {81}},\ \bibinfo {pages} {174411}
  (\bibinfo {year} {2010})}\BibitemShut {NoStop}%
\bibitem [{\citenamefont {Levin}\ and\ \citenamefont {Nave}(2007)}]{TRG}%
  \BibitemOpen
  \bibfield  {author} {\bibinfo {author} {\bibfnamefont {M.}~\bibnamefont
  {Levin}}\ and\ \bibinfo {author} {\bibfnamefont {C.~P.}\ \bibnamefont
  {Nave}},\ }\href {\doibase 10.1103/PhysRevLett.99.120601} {\bibfield
  {journal} {\bibinfo  {journal} {Phys. Rev. Lett.}\ }\textbf {\bibinfo
  {volume} {99}},\ \bibinfo {pages} {120601} (\bibinfo {year}
  {2007})}\BibitemShut {NoStop}%
\bibitem [{\citenamefont {Xie}\ \emph {et~al.}(2009)\citenamefont {Xie},
  \citenamefont {Jiang}, \citenamefont {Chen}, \citenamefont {Weng},\ and\
  \citenamefont {Xiang}}]{SRG}%
  \BibitemOpen
  \bibfield  {author} {\bibinfo {author} {\bibfnamefont {Z.~Y.}\ \bibnamefont
  {Xie}}, \bibinfo {author} {\bibfnamefont {H.~C.}\ \bibnamefont {Jiang}},
  \bibinfo {author} {\bibfnamefont {Q.~N.}\ \bibnamefont {Chen}}, \bibinfo
  {author} {\bibfnamefont {Z.~Y.}\ \bibnamefont {Weng}}, \ and\ \bibinfo
  {author} {\bibfnamefont {T.}~\bibnamefont {Xiang}},\ }\href {\doibase
  10.1103/PhysRevLett.103.160601} {\bibfield  {journal} {\bibinfo  {journal}
  {Phys. Rev. Lett.}\ }\textbf {\bibinfo {volume} {103}},\ \bibinfo {pages}
  {160601} (\bibinfo {year} {2009})}\BibitemShut {NoStop}%
\bibitem [{\citenamefont {Zhao}\ \emph {et~al.}(2016)\citenamefont {Zhao},
  \citenamefont {Xie}, \citenamefont {Xiang},\ and\ \citenamefont
  {Imada}}]{SRGfinite}%
  \BibitemOpen
  \bibfield  {author} {\bibinfo {author} {\bibfnamefont {H.-H.}\ \bibnamefont
  {Zhao}}, \bibinfo {author} {\bibfnamefont {Z.-Y.}\ \bibnamefont {Xie}},
  \bibinfo {author} {\bibfnamefont {T.}~\bibnamefont {Xiang}}, \ and\ \bibinfo
  {author} {\bibfnamefont {M.}~\bibnamefont {Imada}},\ }\href {\doibase
  10.1103/PhysRevB.93.125115} {\bibfield  {journal} {\bibinfo  {journal} {Phys.
  Rev. B}\ }\textbf {\bibinfo {volume} {93}},\ \bibinfo {pages} {125115}
  (\bibinfo {year} {2016})}\BibitemShut {NoStop}%
\bibitem [{\citenamefont {Xie}\ \emph {et~al.}(2012)\citenamefont {Xie},
  \citenamefont {Chen}, \citenamefont {Qin}, \citenamefont {Zhu}, \citenamefont
  {Yang},\ and\ \citenamefont {Xiang}}]{HOTRG}%
  \BibitemOpen
  \bibfield  {author} {\bibinfo {author} {\bibfnamefont {Z.~Y.}\ \bibnamefont
  {Xie}}, \bibinfo {author} {\bibfnamefont {J.}~\bibnamefont {Chen}}, \bibinfo
  {author} {\bibfnamefont {M.~P.}\ \bibnamefont {Qin}}, \bibinfo {author}
  {\bibfnamefont {J.~W.}\ \bibnamefont {Zhu}}, \bibinfo {author} {\bibfnamefont
  {L.~P.}\ \bibnamefont {Yang}}, \ and\ \bibinfo {author} {\bibfnamefont
  {T.}~\bibnamefont {Xiang}},\ }\href {\doibase 10.1103/PhysRevB.86.045139}
  {\bibfield  {journal} {\bibinfo  {journal} {Phys. Rev. B}\ }\textbf {\bibinfo
  {volume} {86}},\ \bibinfo {pages} {045139} (\bibinfo {year}
  {2012})}\BibitemShut {NoStop}%
\bibitem [{\citenamefont {Evenbly}\ and\ \citenamefont {Vidal}(2015)}]{TNR}%
  \BibitemOpen
  \bibfield  {author} {\bibinfo {author} {\bibfnamefont {G.}~\bibnamefont
  {Evenbly}}\ and\ \bibinfo {author} {\bibfnamefont {G.}~\bibnamefont
  {Vidal}},\ }\href {\doibase 10.1103/PhysRevLett.115.180405} {\bibfield
  {journal} {\bibinfo  {journal} {Phys. Rev. Lett.}\ }\textbf {\bibinfo
  {volume} {115}},\ \bibinfo {pages} {180405} (\bibinfo {year}
  {2015})}\BibitemShut {NoStop}%
\bibitem [{\citenamefont {Yang}\ \emph {et~al.}(2017)\citenamefont {Yang},
  \citenamefont {Gu},\ and\ \citenamefont {Wen}}]{loop-TNR}%
  \BibitemOpen
  \bibfield  {author} {\bibinfo {author} {\bibfnamefont {S.}~\bibnamefont
  {Yang}}, \bibinfo {author} {\bibfnamefont {Z.-C.}\ \bibnamefont {Gu}}, \ and\
  \bibinfo {author} {\bibfnamefont {X.-G.}\ \bibnamefont {Wen}},\ }\href
  {\doibase 10.1103/PhysRevLett.118.110504} {\bibfield  {journal} {\bibinfo
  {journal} {Phys. Rev. Lett.}\ }\textbf {\bibinfo {volume} {118}},\ \bibinfo
  {pages} {110504} (\bibinfo {year} {2017})}\BibitemShut {NoStop}%
\bibitem [{\citenamefont {Arnoldi}(1951)}]{Arnoldi}%
  \BibitemOpen
  \bibfield  {author} {\bibinfo {author} {\bibfnamefont {W.~E.}\ \bibnamefont
  {Arnoldi}},\ }\href {\doibase 10.1090/qam/42792} {\bibfield  {journal}
  {\bibinfo  {journal} {Quart. Appl. Math.}\ }\textbf {\bibinfo {volume} {9}},\
  \bibinfo {pages} {17} (\bibinfo {year} {1951})}\BibitemShut {NoStop}%
\bibitem [{\citenamefont {Halko}\ \emph {et~al.}(2011)\citenamefont {Halko},
  \citenamefont {Martinsson},\ and\ \citenamefont {Tropp}}]{RSVD}%
  \BibitemOpen
  \bibfield  {author} {\bibinfo {author} {\bibfnamefont {N.}~\bibnamefont
  {Halko}}, \bibinfo {author} {\bibfnamefont {P.~G.}\ \bibnamefont
  {Martinsson}}, \ and\ \bibinfo {author} {\bibfnamefont {J.~A.}\ \bibnamefont
  {Tropp}},\ }\href {\doibase 10.1137/090771806} {\bibfield  {journal}
  {\bibinfo  {journal} {SIAM Review}\ }\textbf {\bibinfo {volume} {53}},\
  \bibinfo {pages} {217} (\bibinfo {year} {2011})}\BibitemShut {NoStop}%
\bibitem [{\citenamefont {Tamascelli}\ \emph {et~al.}(2015)\citenamefont
  {Tamascelli}, \citenamefont {Rosenbach},\ and\ \citenamefont
  {Plenio}}]{TEBD_RSVD}%
  \BibitemOpen
  \bibfield  {author} {\bibinfo {author} {\bibfnamefont {D.}~\bibnamefont
  {Tamascelli}}, \bibinfo {author} {\bibfnamefont {R.}~\bibnamefont
  {Rosenbach}}, \ and\ \bibinfo {author} {\bibfnamefont {M.~B.}\ \bibnamefont
  {Plenio}},\ }\href {\doibase 10.1103/PhysRevE.91.063306} {\bibfield
  {journal} {\bibinfo  {journal} {Phys. Rev. E}\ }\textbf {\bibinfo {volume}
  {91}},\ \bibinfo {pages} {063306} (\bibinfo {year} {2015})}\BibitemShut
  {NoStop}%
\bibitem [{\citenamefont {Kohn}\ \emph {et~al.}(2018)\citenamefont {Kohn},
  \citenamefont {Tschirsich}, \citenamefont {Keck}, \citenamefont {Plenio},
  \citenamefont {Tamascelli},\ and\ \citenamefont {Montangero}}]{TEBD_RSVD2}%
  \BibitemOpen
  \bibfield  {author} {\bibinfo {author} {\bibfnamefont {L.}~\bibnamefont
  {Kohn}}, \bibinfo {author} {\bibfnamefont {F.}~\bibnamefont {Tschirsich}},
  \bibinfo {author} {\bibfnamefont {M.}~\bibnamefont {Keck}}, \bibinfo {author}
  {\bibfnamefont {M.~B.}\ \bibnamefont {Plenio}}, \bibinfo {author}
  {\bibfnamefont {D.}~\bibnamefont {Tamascelli}}, \ and\ \bibinfo {author}
  {\bibfnamefont {S.}~\bibnamefont {Montangero}},\ }\href {\doibase
  10.1103/PhysRevE.97.013301} {\bibfield  {journal} {\bibinfo  {journal} {Phys.
  Rev. E}\ }\textbf {\bibinfo {volume} {97}},\ \bibinfo {pages} {013301}
  (\bibinfo {year} {2018})}\BibitemShut {NoStop}%
\bibitem [{\citenamefont {Vidal}(2003)}]{TEBD2003}%
  \BibitemOpen
  \bibfield  {author} {\bibinfo {author} {\bibfnamefont {G.}~\bibnamefont
  {Vidal}},\ }\href {\doibase 10.1103/PhysRevLett.91.147902} {\bibfield
  {journal} {\bibinfo  {journal} {Phys. Rev. Lett.}\ }\textbf {\bibinfo
  {volume} {91}},\ \bibinfo {pages} {147902} (\bibinfo {year}
  {2003})}\BibitemShut {NoStop}%
\bibitem [{\citenamefont {Vidal}(2004)}]{TEBD2004}%
  \BibitemOpen
  \bibfield  {author} {\bibinfo {author} {\bibfnamefont {G.}~\bibnamefont
  {Vidal}},\ }\href {\doibase 10.1103/PhysRevLett.93.040502} {\bibfield
  {journal} {\bibinfo  {journal} {Phys. Rev. Lett.}\ }\textbf {\bibinfo
  {volume} {93}},\ \bibinfo {pages} {040502} (\bibinfo {year}
  {2004})}\BibitemShut {NoStop}%
\bibitem [{\citenamefont {Yu}\ \emph {et~al.}(2014)\citenamefont {Yu},
  \citenamefont {Xie}, \citenamefont {Meurice}, \citenamefont {Liu},
  \citenamefont {Denbleyker}, \citenamefont {Zou}, \citenamefont {Qin},
  \citenamefont {Chen},\ and\ \citenamefont {Xiang}}]{HOTRG_XY}%
  \BibitemOpen
  \bibfield  {author} {\bibinfo {author} {\bibfnamefont {J.~F.}\ \bibnamefont
  {Yu}}, \bibinfo {author} {\bibfnamefont {Z.~Y.}\ \bibnamefont {Xie}},
  \bibinfo {author} {\bibfnamefont {Y.}~\bibnamefont {Meurice}}, \bibinfo
  {author} {\bibfnamefont {Y.}~\bibnamefont {Liu}}, \bibinfo {author}
  {\bibfnamefont {A.}~\bibnamefont {Denbleyker}}, \bibinfo {author}
  {\bibfnamefont {H.}~\bibnamefont {Zou}}, \bibinfo {author} {\bibfnamefont
  {M.~P.}\ \bibnamefont {Qin}}, \bibinfo {author} {\bibfnamefont
  {J.}~\bibnamefont {Chen}}, \ and\ \bibinfo {author} {\bibfnamefont
  {T.}~\bibnamefont {Xiang}},\ }\href {\doibase 10.1103/PhysRevE.89.013308}
  {\bibfield  {journal} {\bibinfo  {journal} {Phys. Rev. E}\ }\textbf {\bibinfo
  {volume} {89}},\ \bibinfo {pages} {013308} (\bibinfo {year}
  {2014})}\BibitemShut {NoStop}%
\bibitem [{\citenamefont {Pfeifer}\ \emph {et~al.}(2014)\citenamefont
  {Pfeifer}, \citenamefont {Haegeman},\ and\ \citenamefont
  {Verstraete}}]{NCON}%
  \BibitemOpen
  \bibfield  {author} {\bibinfo {author} {\bibfnamefont {R.~N.~C.}\
  \bibnamefont {Pfeifer}}, \bibinfo {author} {\bibfnamefont {J.}~\bibnamefont
  {Haegeman}}, \ and\ \bibinfo {author} {\bibfnamefont {F.}~\bibnamefont
  {Verstraete}},\ }\href {\doibase 10.1103/PhysRevE.90.033315} {\bibfield
  {journal} {\bibinfo  {journal} {Phys. Rev. E}\ }\textbf {\bibinfo {volume}
  {90}},\ \bibinfo {pages} {033315} (\bibinfo {year} {2014})}\BibitemShut
  {NoStop}%
\bibitem [{\citenamefont {van~der Walt}\ \emph {et~al.}(2011)\citenamefont
  {van~der Walt}, \citenamefont {Colbert},\ and\ \citenamefont
  {Varoquaux}}]{NUMPY}%
  \BibitemOpen
  \bibfield  {author} {\bibinfo {author} {\bibfnamefont {S.}~\bibnamefont
  {van~der Walt}}, \bibinfo {author} {\bibfnamefont {S.~C.}\ \bibnamefont
  {Colbert}}, \ and\ \bibinfo {author} {\bibfnamefont {G.}~\bibnamefont
  {Varoquaux}},\ }\href {\doibase 10.1109/MCSE.2011.37} {\bibfield  {journal}
  {\bibinfo  {journal} {Computing in Science \& Engineering}\ }\textbf
  {\bibinfo {volume} {13}},\ \bibinfo {pages} {22} (\bibinfo {year}
  {2011})}\BibitemShut {NoStop}%
\bibitem [{\citenamefont {Jones}\ \emph {et~al.}(01  )\citenamefont {Jones},
  \citenamefont {Oliphant}, \citenamefont {Peterson} \emph {et~al.}}]{SCIPY}%
  \BibitemOpen
  \bibfield  {author} {\bibinfo {author} {\bibfnamefont {E.}~\bibnamefont
  {Jones}}, \bibinfo {author} {\bibfnamefont {T.}~\bibnamefont {Oliphant}},
  \bibinfo {author} {\bibfnamefont {P.}~\bibnamefont {Peterson}},  \emph
  {et~al.},\ }\href {http://www.scipy.org/} {\enquote {\bibinfo {title}
  {{SciPy}: Open source scientific tools for {Python}},}\ } (\bibinfo {year}
  {2001--})\BibitemShut {NoStop}%
\bibitem [{\citenamefont {Anderson}\ \emph {et~al.}(1999)\citenamefont
  {Anderson}, \citenamefont {Bai}, \citenamefont {Bischof}, \citenamefont
  {Blackford}, \citenamefont {Demmel}, \citenamefont {Dongarra}, \citenamefont
  {Du~Croz}, \citenamefont {Greenbaum}, \citenamefont {Hammarling},
  \citenamefont {McKenney},\ and\ \citenamefont {Sorensen}}]{LAPACK}%
  \BibitemOpen
  \bibfield  {author} {\bibinfo {author} {\bibfnamefont {E.}~\bibnamefont
  {Anderson}}, \bibinfo {author} {\bibfnamefont {Z.}~\bibnamefont {Bai}},
  \bibinfo {author} {\bibfnamefont {C.}~\bibnamefont {Bischof}}, \bibinfo
  {author} {\bibfnamefont {S.}~\bibnamefont {Blackford}}, \bibinfo {author}
  {\bibfnamefont {J.}~\bibnamefont {Demmel}}, \bibinfo {author} {\bibfnamefont
  {J.}~\bibnamefont {Dongarra}}, \bibinfo {author} {\bibfnamefont
  {J.}~\bibnamefont {Du~Croz}}, \bibinfo {author} {\bibfnamefont
  {A.}~\bibnamefont {Greenbaum}}, \bibinfo {author} {\bibfnamefont
  {S.}~\bibnamefont {Hammarling}}, \bibinfo {author} {\bibfnamefont
  {A.}~\bibnamefont {McKenney}}, \ and\ \bibinfo {author} {\bibfnamefont
  {D.}~\bibnamefont {Sorensen}},\ }\href@noop {} {\emph {\bibinfo {title}
  {{LAPACK} Users' Guide}}},\ \bibinfo {edition} {3rd}\ ed.\ (\bibinfo
  {publisher} {Society for Industrial and Applied Mathematics},\ \bibinfo
  {address} {Philadelphia, PA},\ \bibinfo {year} {1999})\BibitemShut {NoStop}%
\bibitem [{\citenamefont {Lehoucq}\ \emph {et~al.}(1998)\citenamefont
  {Lehoucq}, \citenamefont {Sorensen},\ and\ \citenamefont {Yang}}]{ARPACK}%
  \BibitemOpen
  \bibfield  {author} {\bibinfo {author} {\bibfnamefont {R.}~\bibnamefont
  {Lehoucq}}, \bibinfo {author} {\bibfnamefont {D.}~\bibnamefont {Sorensen}}, \
  and\ \bibinfo {author} {\bibfnamefont {C.}~\bibnamefont {Yang}},\ }\href
  {\doibase 10.1137/1.9780898719628} {\emph {\bibinfo {title} {ARPACK Users'
  Guide}}}\ (\bibinfo  {publisher} {Society for Industrial and Applied
  Mathematics},\ \bibinfo {address} {Philadelphia, PA},\ \bibinfo {year}
  {1998})\BibitemShut {NoStop}%
\end{thebibliography}%

\end{document}